\documentstyle[epsf]{mn}
\title[The origin and formation of cuspy density profiles]
{The origin and formation of cuspy density profiles through violent relaxation
of stellar systems}

\author[S.~Hozumi, A.~Burkert and T.~Fujiwara]
{S.~Hozumi,$^1$\thanks{E-mail: hozumi@sue.shiga-u.ac.jp (SH);
burkert@mpia-hd.mpg.de (AB); fujiwara@kcua.ac.jp (TF)}
 A.~Burkert$^{2\,{\mbox{\LARGE$\star$}}}$ and
 T.~Fujiwara$^{3\,{\mbox{\LARGE$\star$}}}$\\
$^1$Faculty of Education, Shiga University, 2-5-1 Hiratsu,
Otsu, Shiga 520-0862, Japan\\
$^2$Max-Planck-Institut f\"ur Astronomie, K\"onigstuhl 17, D-69117
Heidelberg, Germany\\
$^3$Kyoto City University of Arts, Nishikyo-ku, Kyoto 610-1197, Japan}

\date{Accepted 1999 xxx xx.
      Received 1999 January 4}

\pagerange{\pageref{firstpage}--\pageref{lastpage}}
\pubyear{1999}

\begin{document}

\maketitle

\label{firstpage}

\begin{abstract}
It is shown that the cuspy density distributions observed in the cores
of elliptical galaxies can be realized by dissipationless gravitational
collapse.  The initial models consist of power-law density spheres such
as $\rho\propto r^{-1}$ with anisotropic velocity dispersions.  Collapse
simulations are carried out by integrating the collisionless Boltzmann
equation directly, on the assumption of spherical symmetry.  From the
results obtained, the extent of constant density cores, formed through
violent relaxation, decreases as the velocity anisotropy increases
radially, and practically disappears for extremely radially anisotropic
models.  As a result, the relaxed density distributions become more cuspy
with increasing radial velocity anisotropy.  It is thus concluded that
the velocity anisotropy could be a key ingredient for the formation of
density cusps in a dissipationless collapse picture.  The velocity
dispersions increase with radius in the cores according to the nearly
power-law density distributions.  The power-law index, $n$, of the density
profiles, defined as $\rho\propto r^{-n}$, changes from $n\approx 2.1$ at
intermediate radii, to a shallower power than $n\approx 2.1$ toward the
centre.  This density bend can be explained from our postulated {\it local}
phase-space constraint that the phase-space density accessible to the
relaxed state is determined at each radius by the maximum phase-space
density of the initial state.
\end{abstract}


\section{Introduction}
Recent {\it Hubble Space Telescope} observations (Crane et al.\ 1993;
Ferrarese et al.\ 1994; Jaffe et al.\ 1994; Lauer et al.\ 1995; Byun et
al.\ 1996; Gebhardt et al.\ 1996; Faber et al.\ 1997) have revealed that
elliptical galaxies do not have constant density cores but have cusps
which continue toward the centre until the resolution limit.  In addition,
it has been found that the density profiles of the cores are divided into
two types: one is a shallow cuspy core represented by $\rho\propto r^{-n}$
with $0.5\la n\la 1$ for bright elliptical galaxies, and the other is
a steep cuspy core described by approximately $\rho\propto r^{-2}$ for
faint ones (Merritt \& Fridman 1996).  Several cosmological simulations,
based on a standard cold dark matter scenario, have demonstrated that
cuspy density profiles with $\rho\propto r^{-1}$ might be a natural
end-product of hierarchical clustering (Dubinski \& Carlberg 1991;
Navarro, Frenk \& White 1996, 1997).  Recently, Fukushige \& Makino
(1997) have pointed out from their high resolution simulation with
about $8\times 10^5$ particles that the resulting central density
profiles of dark matter haloes are steeper than $\rho\propto r^{-1}$
in a hierarchical clustering scenario, and that merging processes
among clumps with different binding energies would play a significant
role in the formation of such steep density cusps.  Moore et al.\ (1998)
have also demonstrated that very high resolution simulations with nearly
$3\times 10^6$ particles yield steep inner density profiles with $\rho
\propto r^{-1.4}$ (see however Kravtsov et al.\ 1998).  On the other hand,
shallower density cusps could be formed through merging between galaxies
which individually have a massive central black hole, as shown by Makino
\& Ebisuzaki (1996).  For this kind of merging, Nakano \& Makino (1999)
have explained why a galaxy that has swallowed a massive black hole
leaves a shallow cusp in the centre.  However, the physical mechanism to
generate steep cusps still remains unsolved.

From the simulations mentioned above, we might conclude that clumps are
needed to produce steep cusps while merging between galaxies with massive
central black holes is required to generate shallow cusps.  However, it
might be too early to abandon a simple dissipationless collapse picture,
which includes neither clumpiness nor massive central black holes, for
the formation of cusps.  We know that the collapse of uniform-density
spheres results in large constant density cores (van Albada 1982; Fujiwara
1983), while little is known about the details of the collapse of
power-law density spheres.  Of course, there are cases for the collapse
of non-uniform density spheres.  For example, Hozumi, Fujiwara \& Kan-ya
(1996) have carried out collapse simulations of power-law density spheres
only in the initial contraction phases in order to demonstrate the
predominance of the growth of the tangential velocity dispersion over
that of the radial velocity dispersion until the most contracting phase.
In addition, Londrillo, Messina \& Stiavelli (1991) examined the collapse
of generalized Plummer models; their models have large cores, and steep
density haloes with $\rho\propto r^{-5}$ or $\rho\propto r^{-7}$, so that
the collapse behaviour over most parts of the system is considered similar
to that of a uniform-density sphere.  Burkert (1990) paid attention to
degenerate cores arising from the dissipationless collapse of spherical
stellar systems with different initial density gradients and virial ratios.
In another instance, Cannizzo \& Hollister (1992) studied the relation
between the power-law index of initial density distributions and the final
axis ratios of the systems.  However, the structures at sufficiently small
radii where the density cusps are observed in real galaxies were not a
main concern in these examples.

Although many dissipationless collapse simulations have been carried out
so far, there has been no sufficiently high resolution to determine fine
structures at very small radii, such as density cusps. In addition, the
systems in numerical studies were assumed to have isotropic velocity
distributions at the beginning.  That is, the initial models for
dissipationless collapse simulations have never included anisotropic
velocity distributions, despite the fact that stellar systems can support
velocity anisotropy unlike gaseous systems.

In this paper, we examine the influence of initial velocity anisotropy on
the relaxed density profiles in the core, assigning enough resolution to
analyse the innermost region, and show that radially anisotropic models
could produce different cuspy density profiles.  In Section 2, we describe
the initial models in which anisotropic velocity distributions are taken
into account, and the numerical method.  In Section 3, the results of
collapse simulations are presented.  In Section 4, we explain the resulting
density profiles from a viewpoint that the maximum phase-space density
accessible to the relaxed system is determined at each radius by that of
the initial system.  Conclusions are given in Section 5.

\section{Models and Method}
We carry out collapse simulations of power-law density spheres on the
assumption of spherical symmetry.  The initial virial ratios, $\eta\equiv
2T/|W|$, used are $\eta=0.1$ and $\eta=10^{-1.5}$, where $T$ is the kinetic
energy, and $W$ is the potential energy.  According to existing simulations
(Polyachenko 1981, 1992; Merritt \& Aguilar 1985; Min \& Choi 1989; Aguilar
\& Merritt 1990; Londrillo et al.\ 1991; Cannizzo \& Hollister 1992; Udry
1993) like those studied here, our models would be affected by the radial
orbit instability if spherical symmetry were not assumed, because of the
small initial virial ratios.  However, Aguilar \& Merritt (1990) have shown
that the origin of the radial orbit instability is different from that of
the universal density profile observed in elliptical galaxies, that is,
the de Vaucouleurs $R^{1/4}$ law in projection (see also Burkert 1993).
Therefore, we are allowed to assume spherical symmetry as long as we focus
only on the spherically averaged density profiles and related physical
quantities of collapsed objects.

The velocity anisotropy is introduced into the initial models by adopting
the distribution function, $f$, as
\begin{equation}
  f=\frac{\rho_0(r)}{\sqrt 2\pi^{3/2}\sigma_r\sigma_\bot^2}
    \exp\left(-\frac{u^2}{2\sigma_r^2}\right)
    \exp\left(-\frac{j^2/r^2}{\sigma_\bot^2}\right),
  \label{dfeq}
\end{equation}
where $\rho_0$ is the density, $\sigma_r$ is the radial velocity
dispersion, and $\sigma_\bot$ is the tangential velocity dispersion with
$r$, $u$, and $j$ being the radius, radial velocity, and angular momentum,
respectively.  We begin with a moderate power-law density profile given by
\begin{equation}
  \rho_0(r)=(M/2\pi R_0^2)r^{-1},
\end{equation}
where $M$ is the total mass and $R_0$ is the radius of the sphere.  We
measure the velocity anisotropy by $\alpha\equiv 2T_r/T_\bot$, where
$T_r$ and $T_\bot$ are the kinetic energies in radial and tangential
motions, respectively.  In our models, $\alpha$ is equivalent to
$2\sigma_r^2/\sigma_\bot^2$, as is found from equation (\ref{dfeq}).
We take $\alpha=0.2, 0.5, 1, 2, 6,$ and 10 for each value of the virial
ratios.

We use Fujiwara's (1983) phase-space solver which integrates the
collisionless Boltzmann equation directly, with the help of a splitting
scheme (Cheng \& Knorr 1976).  Since this method leads to the results that
are not subject to random fluctuations, we can obtain smooth density and
velocity dispersion profiles.  We employ $(N_r, N_u, N_j)=(300, 301, 50)$,
where $N_r, N_u$, and $N_j$ are the numbers of mesh points along the radius,
radial velocity, and angular momentum, respectively.  A logarithmic grid
is used in the radial direction.  An equal interval grid is assigned to
the radial velocity while a quadratic interval is adopted for the angular
momentum.  The units of mass, $M$, and the gravitational constant, $G$,
are taken to be $M=G=1$.  The unit of length is determined from the
relation such that $\eta\times R_0=1$.  This choice of $R_0$ leads to
approximately the same core sizes for uniform-density spheres with
isotropic velocity distributions (Fujiwara 1983).  As a characteristic
time-scale for the collapse, we adopt the free-fall time, defined by
$t_{\rm ff}=\pi\sqrt{R_0^3/8GM}$.  We obtain $t_{\rm ff}=35.1$ for
$\eta=0.1$ and $t_{\rm ff}=197$ for $\eta=10^{-1.5}$ in our system of
units.  The minimum and maximum radii assigned to grid points are set
to be $R_{\rm min}=0.01$ and $R_{\rm max}=40.0$, respectively, for
$\eta=0.1$, and $R_{\rm min}=0.01$ and $R_{\rm max}=49.9$, respectively,
for $\eta=10^{-1.5}$.  Test simulations have shown that resulting density
and velocity dispersion profiles at small radii do not depend on the
replacement of $R_{\rm min}=0.01$ by $R_{\rm min}=0.005$ with more finely
divided mesh points in the angular momentum.

Hozumi \& Hernquist (1995) and Hozumi (1997) traced the orbits of stars
on grid points backward to $t=0$ every suitable time step in order to
avoid the numerical diffusion generated by the repeated interpolation
required by the splitting scheme.  However, we do not adopt this method
in our simulations.  This is because we are interested not in the details
of the phase-space distributions but mainly in low-order moments of the
distribution function such as density and velocity dispersion, in addition
to saving computational time.

\section{Results}
We stopped the simulations at $t=120$ for $\eta=0.1$ and at $t=500$
for $\eta=10^{-1.5}$.  These times correspond to about $3\; t_{\rm ff}$.
They are, nevertheless, sufficiently long that the resulting cores have
relaxed completely.  Since our main concern is the core regions, these
times satisfy our requirements for analysing the relaxed density and
velocity dispersion profiles near those regions.  There were practically
no escapers for $\eta=0.1$, while about 20 per cent of the total mass
with positive energies expanded beyond $R_{\rm max}$ and escaped for
$\eta=10^{-1.5}$.  Consequently, the total energy was conserved to better
than 0.39 per cent for all the models with $\eta=0.1$ except for the model
with $\alpha=0.2$ in which the total energy within $R_{\rm max}$ changed
by 1.8 per cent owning to a small fraction of escapers.  On the other hand,
the total energy within $R_{\rm max}$ changed by at worst 10.7 per cent for
all the models with $\eta=10^{-1.5}$ because of a large fraction of escapers.

In Figs.~1a and 1b, we show the resulting density profiles for $\eta=0.1$
and $\eta=10^{-1.5}$, respectively.  The half-mass radii, $r_{\rm h}$,
for $\eta=0.1$ are about 2.6 to 3.2, while they are about 7.8 to 9.7
for $\eta=10^{-1.5}$.  Thus, our adopted minimum radii correspond to
$R_{\rm min}\sim 0.003\; r_{\rm h}$ for $\eta=0.1$ and $R_{\rm min}\sim
0.001\; r_{\rm h}$ for $\eta=10^{-1.5}$.  If the half-mass radius is
regarded practically as the effective radius, $r_{\rm e}$, of the de
Vaucouleurs law, the minimum radii used are sufficiently small and
correspond to roughly 10 pc, because $r_{\rm e}$ is typically 3 kpc for
bright ellipticals and is smaller for faint ones (Kormendy 1977).
We point out that most galaxies are not well resolved inside to 10 pc
according to the data cited in the paper of Gebhardt et al.\ (1996).

We can see from Fig.~1 that the density distributions at small radii
become more cuspy as $\alpha$ increases, and that their slope becomes
steeper with increasing $\alpha$ for both virial ratios.  As a result,
the core region, specified by a nearly constant density distribution,
decreases with increasing $\alpha$.  In particular, it is to be noticed
that there is no constant density core for the most radially anisotropic
models, that is, the models with $\alpha=10$.  On the other hand, the
density profiles at intermediate radii are well-approximated by $\rho
\propto r^{-2.1}$, regardless of the virial ratios and the velocity
anisotropy.

The relaxed velocity dispersion profiles for $\eta=0.1$ and $\eta=10^{-1.5}$
are shown in Figs.~2a and 2b, respectively.  These figures indicate that the
relaxed velocity distributions no longer remain anisotropic in the inner
region of the cores even though we begin with a model which has a large
velocity anisotropy at all radii.  Thus, it seems very difficult to produce
and maintain velocity anisotropy in the cores only by gravitational collapse.
The isotropic region that roughly corresponds to the core becomes smaller,
as the velocity anisotropy increases.  It is found from Figs.~2a and 2b that
the density of the inner regions necessarily decreases with radius because
the velocity dispersions rise with radius in the cores.  In particular, for
$\alpha=10$, the velocity dispersions in the cores increase almost linearly
with radius in logarithmic scale.  Then, the hydrostatic equation requires
that the corresponding density profiles should follow a power-law
distribution.  In reality, as shown by Fig.~2, the imperfect linearity of
the velocity dispersion against radius in logarithmic scale does not
realize a perfect power-law density distribution.

\section{Discussion}
\subsection{Phase-space constraint}
Collisionless stellar systems suffer the constraint that the phase-space
density accessible to the relaxed state is less than the maximum phase-space
density of the initial state.  On the basis of this phase-space constraint,
we can explain the density bend which occurs around the edge of the cores.

As far as the core generated through violent relaxation is concerned,
the phase-space density does not decrease substantially from the initial
value at least for spherically symmetric systems (Fujiwara 1983; Hozumi
1997).  Our adopted initial virial ratios are small enough to induce
a radial orbit instability, unless spherical symmetry is imposed.
Nevertheless, it may be unlikely that the phase-space density in the
core suffers a substantial decrease after violent relaxation according
to the results of May \& van Albada (1984):  they have shown that the
phase-space density in the core is almost conserved after violent
relaxation even for aspherical collapses, although they examined this
behaviour only for initially homogeneous density distribution models.
The effects of aspherical collapses through the radial orbit instability
are left to be investigated as a future problem.

Here, admitting the non-decreasing nature of the phase-space density in the
core, we postulate a {\it local} phase-space constraint that the maximum
phase-space density in the relaxed system is determined by the initial
conditions at each radius in the core regions for power-law density spheres.
From equation (\ref{dfeq}), the {\it local} maximum phase-space density at
the beginning, $f_{\rm m}$, is defined by
\begin{equation}
  f_{\rm m}(r)\equiv
  \frac{\rho_0(r)}{\sqrt 2\pi^{3/2}\sigma_r\sigma_\bot^2}.
  \label{deffmax}
\end{equation}
If the initial density distribution is given by a power law like $\rho_0
\propto r^{-n}$, the {\it local} maximum phase-space density of the initial
system is represented in terms of the anisotropic parameter, $\alpha$, and
the virial ratio, $\eta$, by
\begin{equation}
  f_{\rm m}(r)=\frac{1}{8\sqrt 2\pi^{5/2}}\frac{(5-2n)^{3/2}}{(3-n)^{1/2}}
     \frac{(\alpha+2)^{3/2}}{\alpha^{1/2}}\eta^{-n}r^{-n}.
  \label{fmaxeq}
\end{equation}
Notice that equation (\ref{fmaxeq}) holds even though $\alpha$ depends on
radius.  The cumulative mass within $r$ is $M(r)=(r/R_0)^{3-n}$.  For the
special case where $\alpha$ is independent of $r$, we find the cumulative
mass with phase-space density greater than $f_{\rm m}$ by solving equation
(\ref{fmaxeq}) with respect to $r$,
\begin{eqnarray}
  M(\ge f_{\rm m}) & =&\frac{1}{[8\sqrt 2\pi^{5/2}]^{(3-n)/n}}
       \frac{(5-2n)^{3(3-n)/(2n)}}{(3-n)^{(3-n)/(2n)}}
    \nonumber\\
    & & \times\frac{(\alpha+2)^{3(3-n)/(2n)}}{\alpha^{(3-n)/(2n)}}
       f_{\rm m}^{-(3-n)/n}.
  \label{mfeq}
\end{eqnarray}
This equation shows that $M(\ge f_{\rm m})$ is independent of $\eta$,
because we have determined $R_0$ from the relation such that $\eta\times
R_0=1$.  Thus, our discussion described below holds whatever value the
virial ratio will be.  From equation (\ref{mfeq}), we obtain the relation
between $M$ and $f_{\rm m}$ as
\begin{equation}
  M(\ge f_{\rm m})=\frac{27}{256\pi^5}%
                   \frac{(\alpha+2)^3}{\alpha}f_{\rm m}^{-2}
  \quad\quad{\rm for}\quad n=1,
  \label{mf1eq}
\end{equation}
and
\begin{equation}
  M(\ge f_{\rm m})=\frac{1}{2^{7/4}\pi^{5/4}}
                 \frac{(\alpha+2)^{3/4}}{\alpha^{1/4}}f_{\rm m}^{-1/2}
  \quad\quad{\rm for}\quad n=2.
  \label{mf2eq}
\end{equation}

Since our initial models have $\rho_0\propto r^{-1}$, equation
(\ref{mf1eq}) predicts the relation $M(\ge f_{\rm m})\propto
f_{\rm m}^{-2}$.  The resulting $M(\ge f_{\rm m})$ after violent
relaxation cannot exceed $M(\ge f_{\rm m})$ of the initial system.  As the
relaxed density distributions show $\rho\propto r^{-2.1}$ at intermediate
radii, regardless of the values of $\alpha$, $M(\ge f_{\rm m})\propto
f_{\rm m}^{-1/2}$ would hold approximately at such radii.  As illustrated
in Fig.~3, for large values of $f_{\rm m}$ there unavoidably emerges a
region in which $M(\ge f_{\rm m})$ for $\rho\propto r^{-2.1}$ becomes
larger than that for $\rho\propto r^{-1}$.  Since large $f_{\rm m}$
correspond to small $r$ as indicated by equation (\ref{fmaxeq}), we find
from Fig.~3 that the density profile with $\rho\propto r^{-2.1}$ cannot
continue down to sufficiently small radii owning to the lack of phase-space
density.  Therefore, the density bend necessarily arises as long as the
initial density distribution is shallower than $\rho_0\propto r^{-2.1}$.

As a piece of evidence that the {\it local} phase-space constraint
is valid, we present in Fig.~4 the plots of phase particles on the
radial mesh points at the final state for $\eta=0.1$ with $\alpha=0.2$,
1, and 10 in the $(f,r)-$plane.  This figure reveals that phase particles
with high phase-space density are not scattered in a wide range of radius.
Thus, at least in the central region, mixing of high phase-space density
particles with low phase-space ones does not occur efficiently through
violent relaxation.  If we pick out the maximum phase-space density at
each radius from that distribution of the phase particles which is shown
in Fig.~4, we can obtain, in a practical sense, the {\it local} maximum
phase-space density, $f_{\rm m}$, in the relaxed state.  Fig.~5 is
depicted in such a way to show the relaxed $f_{\rm m}$ against $r$ near
the central region.  We can see from this figure that the relaxed
$f_{\rm m}$ is well-approximated by $f_{\rm m}\propto r^{-2}$ at $r\ga
0.1$, although the slope of the model with $\alpha=10$ is closer to
$-1.6$ rather than $-2$.  The relation $f_{\rm m}\propto r^{-2}$ is
naturally derived from equation (\ref{fmaxeq}) for the density profile
$\rho\propto r^{-2}$ which is in fact obtained at intermediate radii
in our simulations, though it is precisely $\rho\propto r^{-2.1}$.
Therefore, our explanation for the density bend based on Fig.~3 is
justified qualitatively.

We now turn to the tendency that larger values of $\alpha$ result in more
cuspy density distributions.  When we begin with $\rho_0\propto r^{-1}$,
equation (\ref{mf1eq}) indicates that $M(\ge f_{\rm m})$ is proportional
to $(\alpha+2)^3/\alpha$.  This function has a minimum at $\alpha=1$.
Hence, $M(\ge f_{\rm m})$ increases with increasing $\alpha$ if $\alpha > 1$.
This means that $M(\ge f_{\rm m})$ for $\rho\propto r^{-2}$ intersects with
$M(\ge f_{\rm m})$ for $\rho\propto r^{-1}$ with $\alpha=10$ at a larger
$f_{\rm m}$ than that with $\alpha=1$ (see Fig.~3).  Thus, $\rho\propto
r^{-2.1}$ can continue down to smaller radii with increasing $\alpha$,
provided that the relaxed density profiles at intermediate radii do not
change greatly among the models with different values of $\alpha$, as is
demonstrated by our simulations.  In this way, we can understand that the
large $\alpha$ models produce nearly cuspy density profiles down to
sufficiently small radii.

On the other hand, $M(\ge f_{\rm m})$ decreases with increasing $\alpha$
if $\alpha < 1$.  Thus, if $\alpha < 1$, smaller values of $\alpha$ should
also, in principle, lead to a density bend closer to the centre (see Fig.~3).
In reality, this is not the case, as illustrated in Fig.~1.  This can be
understood from Fig.~6 which shows the fractional mass, $dM(j)/dj$, with
the angular momentum, $j$, for the initial models.  These fractional mass
distributions remain unchanged throughout the evolution because spherical
symmetry is assumed.  Fig.~6 shows that the mass of stars with small values
of the angular momentum decreases as $\alpha\ (<1)$ decreases for $j\la 0.2$.
If the $\rho\propto r^{-2}$ profile could continue down to the very centre,
and if most stars would be on circular orbits, $dM(j)/dj$ would be equal
to $\sqrt{4\pi\rho_*}$, where $\rho_*$ is the reference density at $r=1$.
The values of the imagined $dM(j)/dj$ are 0.45 for $\eta=0.1$ and 0.26 for
$\eta=10^{-1.5}$.  For these values of $dM(j)/dj$, Fig.~6 indicates that
the smaller $\alpha$ models contain less stars with small $j$.  Since the
stars with small values of the angular momentum can pass close to the
centre, the deficiency of small $j$ stars means that the smaller $\alpha$
models cannot populate a sufficient number of stars down to smaller radii.
Therefore, the radius of the density bend moves outward as $\alpha$
decreases if $\alpha < 1$, even though there is sufficient phase-space
density available.  If spherical symmetry were not imposed, some degree of
mixing in angular momentum could be expected (May \& van Albada 1984), and
so, $\rho\propto r^{-2.1}$ might continue down to smaller radii as $\alpha$
decreases when $\alpha < 1$.

We can infer from Figs.~1 and 6 that the models with $\alpha > 1$ would
not suffer the limitation, described above, on the central density
distribution arising from the amount of small angular momentum stars.
This is because we can explain the shift of the density bend toward
smaller radii with increasing $\alpha$ simply from the {\it local}
phase-space constraint like that illustrated in Fig.~3.

We have found that the velocity anisotropy could be effective in the
formation of cuspy density profiles.  However, it is unclear how the
system can acquire such a large radial anisotropic velocity dispersion.
It may be difficult to realize a large velocity anisotropy in each stellar
system before collapse.  In a cosmological situation based on cold dark
matter, clumps are first formed and then collapse to merge into a large
system.  Thus, a system consisting of clumps may have anisotropic velocity
dispersions.  Therefore, it will be important to study the velocity
distribution just after such a system has decoupled from the Hubble flow.

\subsection{Relation between initial and final density powers}
We have found that more radially anisotropic velocity distributions lead
to more cuspy density profiles by gravitational collapse.  Here, we examine
how seriously the initial power index of density distributions influences
the final one.  We carry out collapse simulations of density distributions
$\rho\propto r^{-n}$ with $n$=0.5, 1, and 2, again on the assumption of
spherical symmetry using the collisionless Boltzmann code.  The anisotropic
parameter and virial ratio are chosen to be $\alpha=10$ and $\eta=0.1$,
respectively.  The other numerical parameters are the same as those used in
Section 2.  The final density profiles are shown in Fig.~7.  The half-mass
radii, $r_{\rm h}$, are about 2.3 to 3.2.  We can see that the density
distributions at intermediate radii are well-approximated by $\rho\propto
r^{-2.1}$, and that initially steeper density distributions result in
steeper cuspy density distributions, although the final power-law indices
within the radius of the density bends become always shallower than the
initial ones.  In particular, the density power for the model with $\rho
\propto r^{-2}$ remains almost unchanged after the collapse, so that
practically no density bend is found.  This fact is in good agreement with
the results of Burkert (1990), and proves the validity of the {\it local}
phase-space constraint that we have postulated in the previous subsection,
because the relaxed density distribution with $\rho\propto r^{-2.1}$ can
almost continue toward the very centre without the phase-space constraint
if the initial density distribution is $\rho\propto r^{-2}$, as indicated
by Fig.~3.

Taking into consideration our results obtained here, the steep cusps
with $\rho\propto r^{-2}$ observed in faint ellipticals could originate
from the fluctuation spectrum that realizes density distributions with
$\rho\propto r^{-2}$.  According to observations, low-luminosity
ellipticals that have the steep cusps are isotropic in the sense that
they show $(V/\sigma)^*\approx 1$, where $(V/\sigma)^*$ is the ratio
of the rotation parameter $V/\sigma$ to the value for an isotropic oblate
spheroid flattened by rotation, $V$ is the maximum rotation velocity, and
$\sigma$ is the mean velocity dispersion inside one-half of the effective
radius (Kormendy \& Bender 1996; Faber et al.\ 1997).  In addition, these
ellipticals are rapidly rotating.  Unfortunately, we cannot discuss
$(V/\sigma)^*$ values for our end-products, because our models include
no net rotation.  Therefore, we will need to investigate the effects of
rotation on the steep cusps using three-dimensional collapses in order to
make a detailed comparison with observations.

The resulting power-law indices of the density distributions within the
radius of the density bend correlate with those of the initial density
distributions.  The difference in power-law index of the cusps originates
from a difference in that of initial density distributions.  In cosmological
simulations with high resolution, based on a standard cold dark matter
scenario, the final density distributions show a rather steep density
cusp, such as $\rho\propto r^{-1.4}$ (Fukushige \& Makino 1997; Moore
et al.\ 1998).  Probably, this kind of violent relaxation would not produce
shallow cusps observed in bright ellipticals, so that we might require a
merger between galaxies each of which contains a massive central black
hole (Makino \& Ebisuzaki 1996; Nakano \& Makino 1999).  However, if the
circumstances for the formation of bright ellipticals are somehow very
different from those of faint ellipticals, it is conceivable from our
results that bright ellipticals could have been formed from the
fluctuation spectrum that corresponds to $\rho\propto r^{-n}$ with
$0.5\la n \la 1$ approximately.

\section{Conclusions}
Collapse simulations of power-law density spheres such as $\rho\propto
r^{-1}$ with anisotropic velocity distributions have been performed on
the assumption of spherical symmetry.  We have found that the resulting
density profiles are well-approximated by $\rho\propto r^{-2.1}$ at
intermediate radii, regardless of the velocity anisotropy and the virial
ratio.  However, this universal density profile cannot continue down to
sufficiently small radii, and so, the density bends necessarily.  We have
explained this density bend from the $local$ phase-space constraint that
the phase-space density accessible to the relaxed system is determined
at each radius by the maximum phase-space density of the initial system.

As the radial velocity anisotropy increases, the density profiles of the
end-products after violent relaxation become more cuspy.  In particular,
the extremely radially anisotropic models with $\alpha=10$ have practically
no flat cores.  Thus, our results imply that the velocity anisotropy could
play a substantial role in the formation of the density cusps observed in
elliptical galaxies as long as we rely on a simple dissipationless collapse
picture.  In reality, it will be difficult to produce a large velocity
anisotropy in individual systems.  Concerning this problem, however, we
point out that a system composed of clumps, as can be envisaged in the
standard cold dark matter scenario, might have a large velocity anisotropy
when it has decoupled from the Hubble expansion.  In addition, we have
examined the relation between the initial and final power-law indices of
the density profiles by performing collapse simulations for $\rho\propto
r^{-n}$ models with $n$=0.5, 1, and 2.  In these cases, the resulting
density profiles at intermediate radii are also well-approximated by
$\rho\propto r^{-2.1}$, regardless of the initial density power indices.
The results show that initially steeper density profiles result in steeper
cuspy density profiles.  Therefore, the difference in the power index of
the cusps between faint and bright elliptical galaxies might arise from
the difference in the power index of initial density profiles.  In
particular, the steep cusps with $\rho\propto r^{-2}$ observed in faint
ellipticals could result from initial density distributions with $\rho
\propto r^{-2}$, because such density distributions can avoid the
{\it local} phase-space constraint.

As a by-product, we have also found that the resulting cores no longer
show velocity anisotropy even though we start with a model which has a
large velocity anisotropy at all radii.  This implies that simple
gravitational collapse could have difficulty in generating and maintaining
velocity anisotropy in the cores.

Our results shown here are obtained on the assumption of spherical
symmetry.  If aspherical collapses are allowed, the radial orbit
instability that arises from small initial virial ratios might affect
the density distributions of end-products.  The study of three-dimensional
collapses is in progress.

\section*{Acknowledgments}
We are grateful to Drs.\ J.\ Makino and T.\ Fukushige for enlightening
discussion.  We are also grateful to the anonymous referee for valuable
comments on the manuscript.  SH thanks Max-Planck-Institut f\"ur Astronomie
in Heidelberg for its hospitality during this research.  This work was
supported in part by the Grant-in-Aid for Scientific Research from the
Ministry of Education, Science, Sports and Culture of Japan (09740172).

\bsp

\label{lastpage}

\clearpage

\newpage

\begin{figure*}
\vspace*{5cm}
\epsfxsize=14cm
\epsfbox{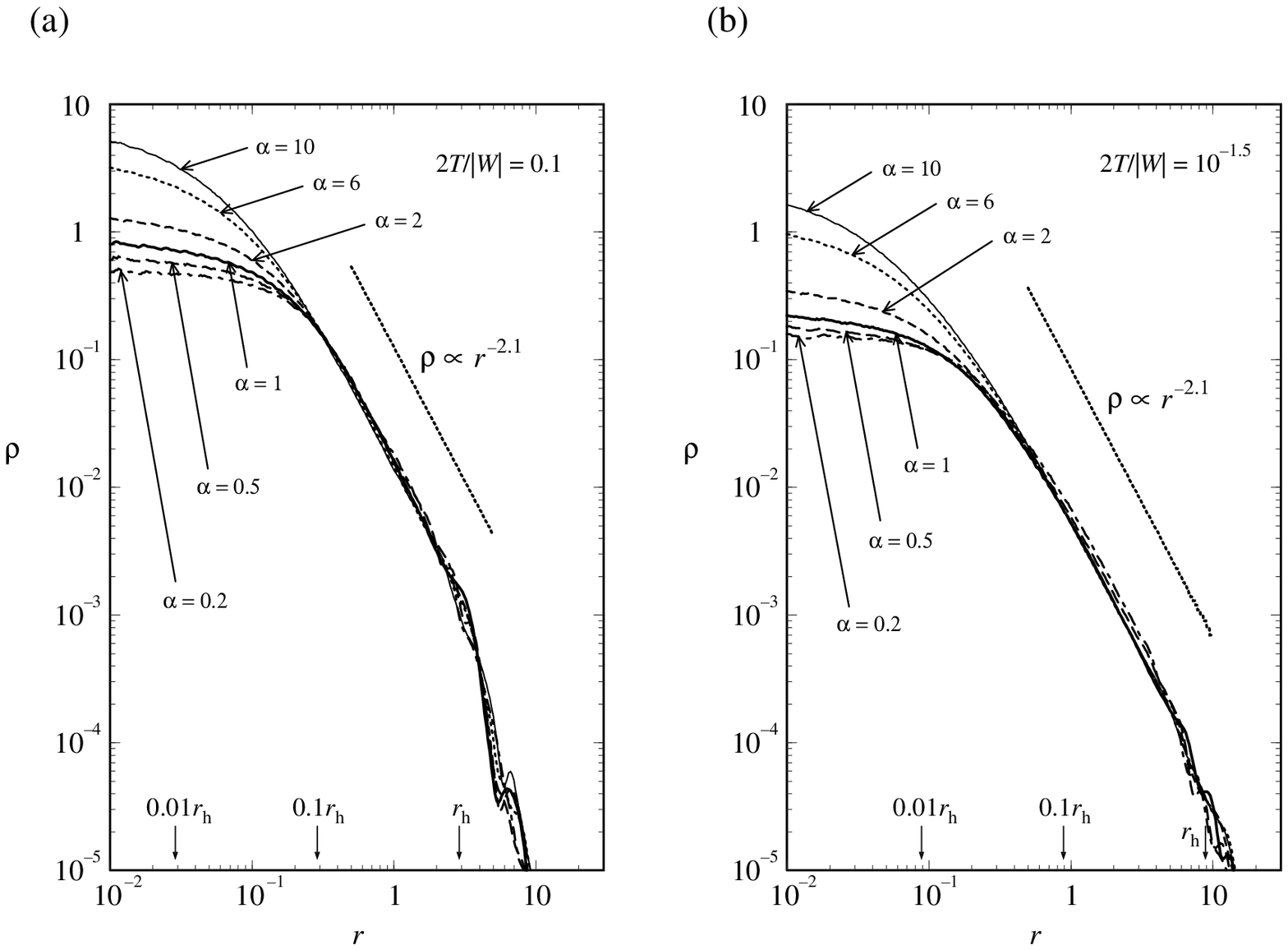}
\caption{
Relaxed density profiles for the models with (a) $2T/|W|=0.1$ and
(b) $2T/|W|=10^{-1.5}$.  The straight dotted lines show the density profiles
with $\rho\propto r^{-2.1}$.  The half-mass radii, $r_{\rm h}$, for
$2T/|W|=0.1$ are about 2.6 to 3.2, while they are about 7.8 to 9.7 for
$2T/|W|=10^{-1.5}$.
}
\label{fig1}
\end{figure*}

\newpage

\begin{figure*}
\epsfxsize=15cm
\epsfbox{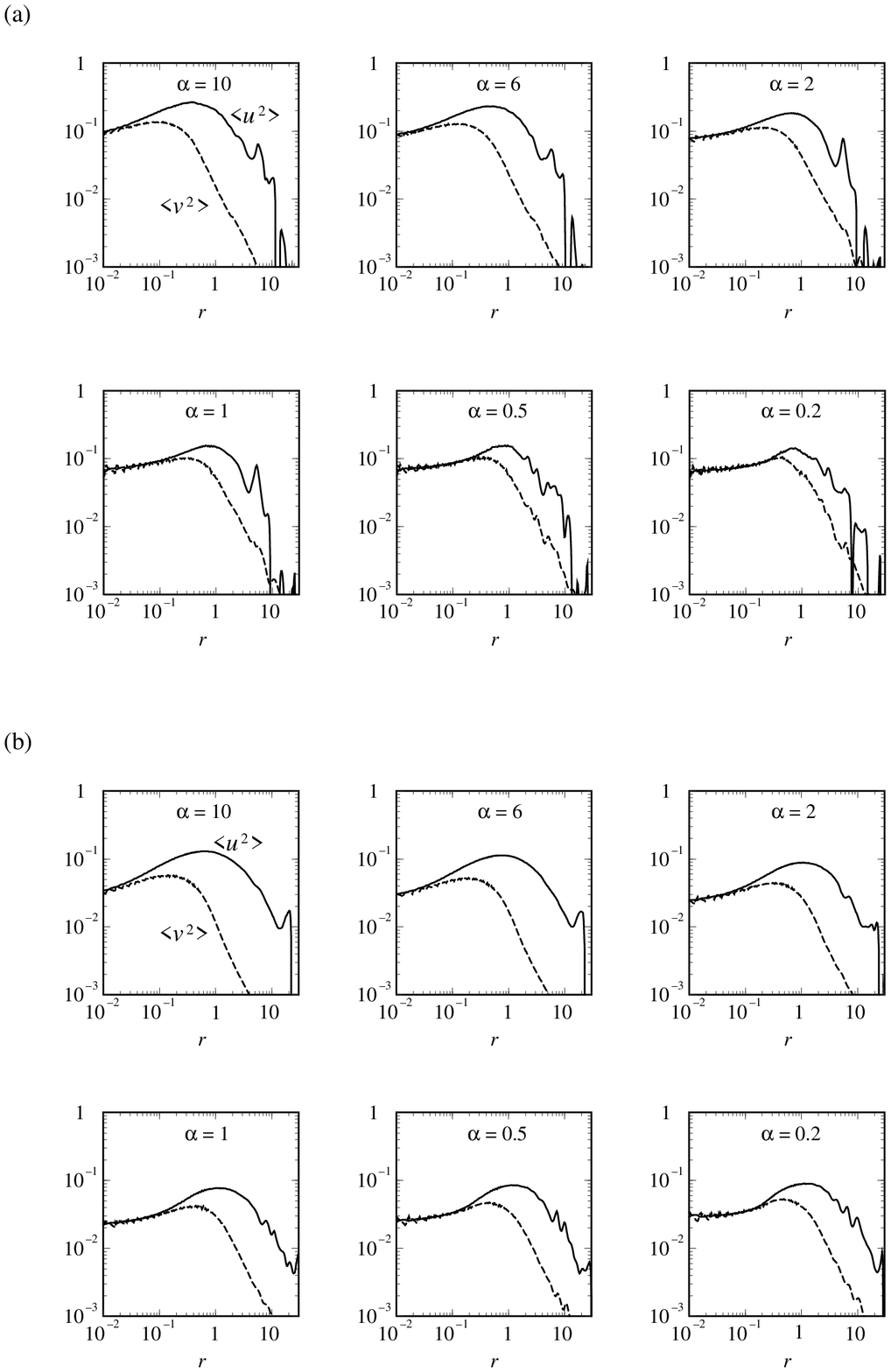}
\caption{
Relaxed velocity dispersion profiles for the models with (a)
$2T/|W|=0.1$ and (b) $2T/|W|=10^{-1.5}$.  The solid lines denote the square
of the radial velocity dispersion, and the dashed lines show that of the
tangential velocity dispersion.
}
\label{fig2}
\end{figure*}

\newpage

\begin{figure*}
\vspace*{5cm}
\epsfxsize=13cm
\epsfbox{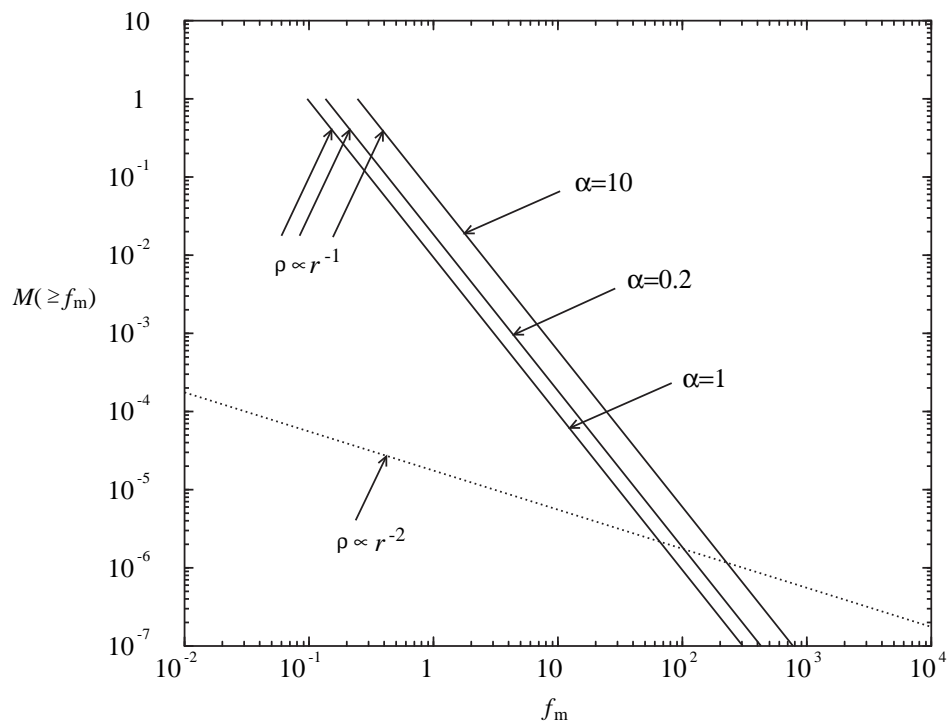}
\caption{
Cumulative mass, $M(\ge f_{\rm m})$, with phase-space density
greater than $f_{\rm m}$ against the maximum phase-space density of the
initial models at each radius, $f_{\rm m}$.  The solid lines denote the
models with $\alpha=0.2, 1,$ and $10$ for the initial density distributions,
$\rho\propto r^{-1}$.  The dotted line represents the mass distribution
for $\rho\propto r^{-2}$.  Notice that the mass distribution in phase-space
density is independent of the initial virial ratio because the initial size
of the sphere, $R_0$, is set to be a reciprocal of the virial ratio.  In
this plot, large $f_{\rm m}$ correspond to small $r$.
}
\label{fig3}
\end{figure*}

\newpage

\begin{figure*}
\vspace*{5cm}
\epsfxsize=17cm
\epsfbox{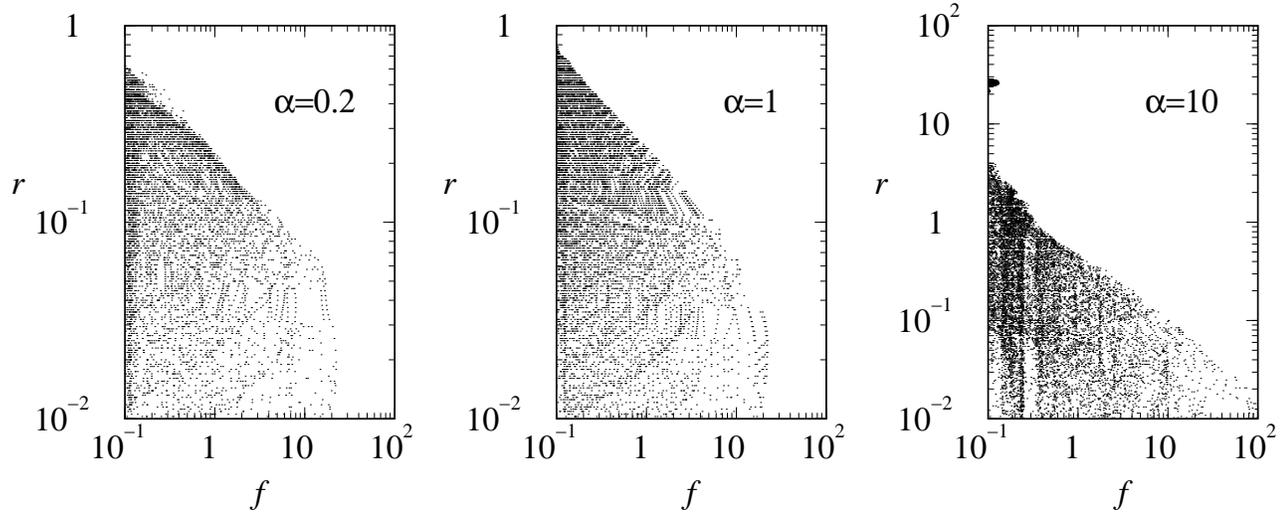}
\caption{
Plots of phase particles on the radial mesh points at the final state
in the $(f,r)-$plane for $\eta=0.1$ with $\alpha=0.1$, 1, and 10, where
$f$ represents the phase-space density of the particles, and $r$ denotes
their radial position.}
\label{fig4}
\end{figure*}

\newpage

\begin{figure*}
\vspace*{5cm}
\epsfbox{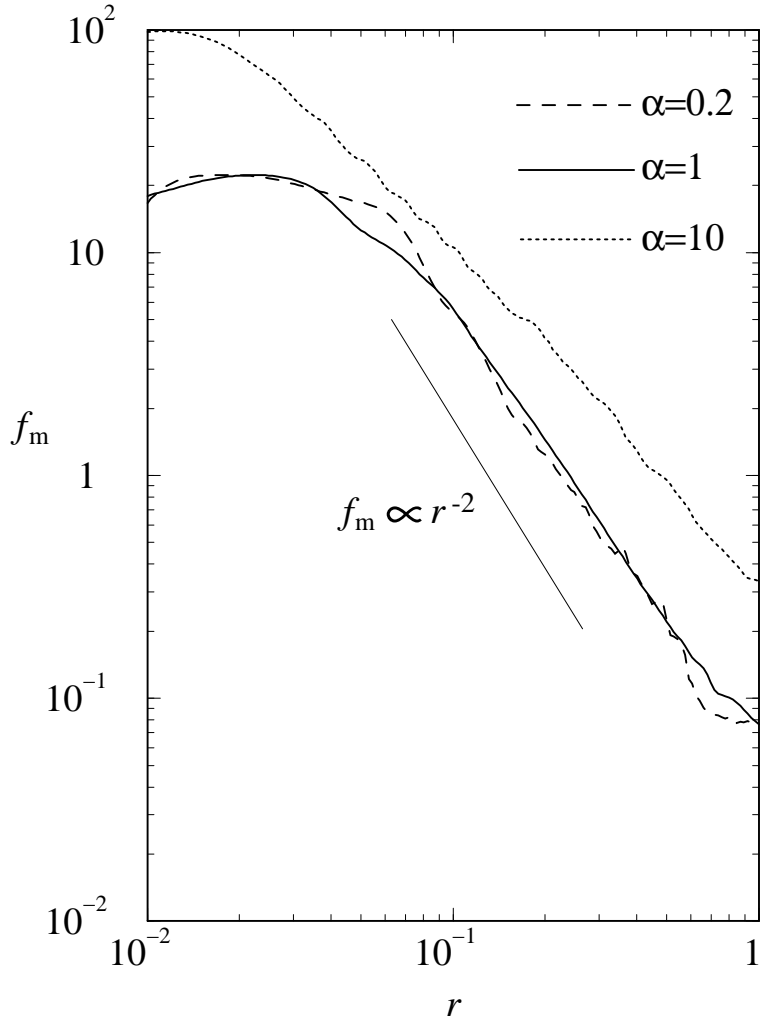}
\caption{
{\it Local} maximum phase-space density in the relaxed state, $f_{\rm m}$,
for $\eta=0.1$ with $\alpha=0.1$, 1, and 10.}
\label{fig5}
\end{figure*}

\newpage

\begin{figure*}
\vspace*{5cm}
\epsfxsize=13cm
\epsfbox{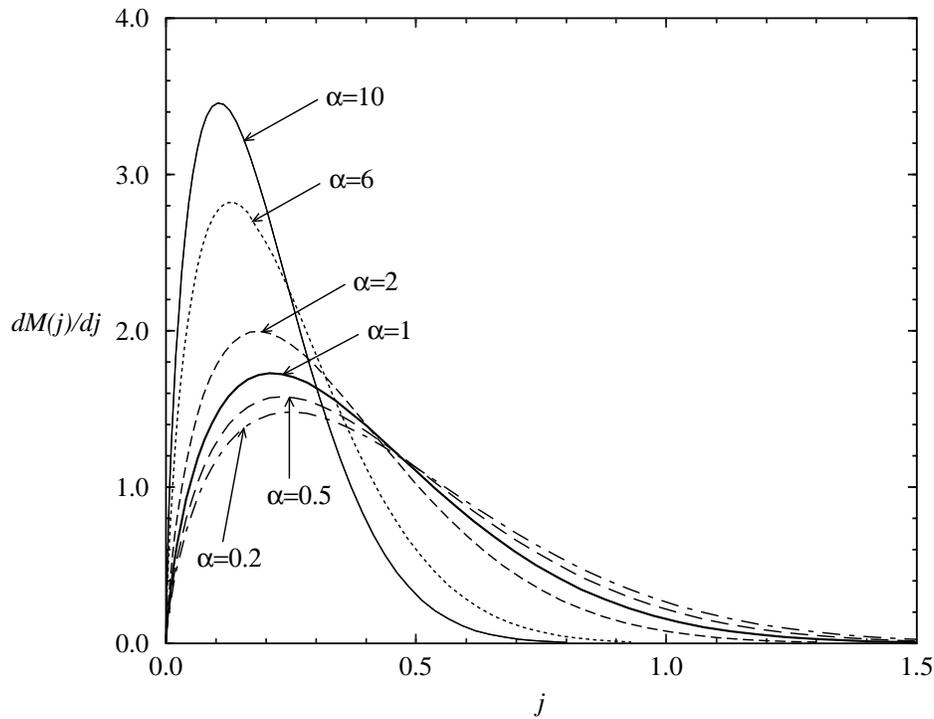}
\caption{
Fractional mass distribution, $dM(j)/dj$, against the angular
momentum, $j$, for the initial models.  Notice that the mass distribution
in angular momentum is independent of the initial virial ratio because the
initial size of the sphere, $R_0$, is set to be a reciprocal of the virial
ratio.
}
\label{fig6}
\end{figure*}

\newpage

\begin{figure*}
\vspace*{5cm}
\epsfbox{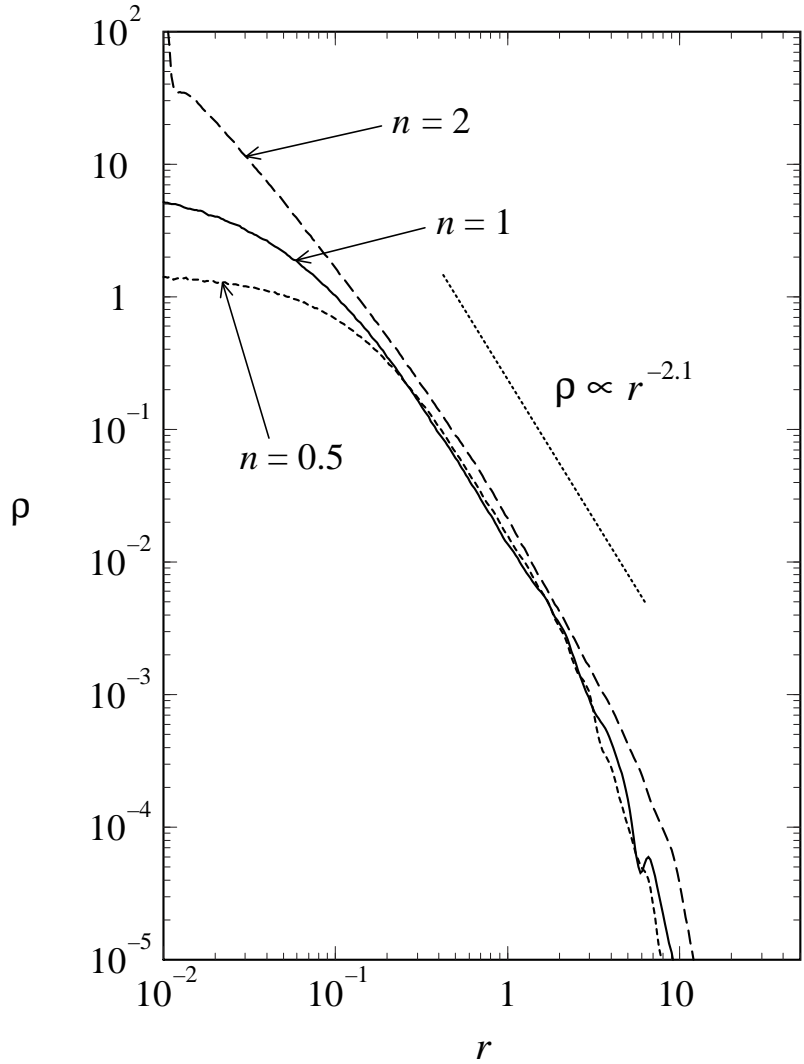}
\caption{
Relaxed density profiles for the power-law density spheres of $\rho
\propto r^{-n}$ with $n$=0.5, 1, and 2.  The virial ratio and the anisotropic
parameter are set to be $\eta=0.1$ and $\alpha=10$, respectively.  The
straight dotted line shows the density profile with $\rho\propto r^{-2.1}$.
}
\label{fig7}
\end{figure*}

\end{document}